\documentclass{article}
\usepackage{authblk}
\usepackage{graphicx}
\usepackage{hyperref}
\usepackage{geometry}
\begin{document}

\renewcommand{\thefootnote}{*}

\title{High-sensitivity low-noise photodetector using large-area silicon photomultiplier}

\author[1]{Takahiko~Masuda\footnote{These authors contributed equally to this work.}}
\author[1]{Ayami~Hiramoto$^*$}
\author[2]{Daniel~G.~Ang}
\author[2]{Cole~Meisenhelder}
\author[3]{Cristian~D.~Panda}
\author[1]{Noboru~Sasao}
\author[1]{Satoshi~Uetake}
\author[2,4]{Xing~Wu}
\author[4]{David~P.~DeMille}
\author[1,2]{John~M.~Doyle}
\author[5]{Gerald~Gabrielse}
\author[1]{Koji~Yoshimura}

\affil[1]{Research Institute for Interdisciplinary Science, Okayama University, Okayama 700-8530, Japan}
\affil[2]{Department of Physics, Harvard University, Cambridge, Massachusetts 02138, USA}
\affil[3]{Department of Physics, University of California, Berkeley, California 94720, USA}
\affil[4]{James Franck Institute and Department of Physics, University of Chicago, Chicago, Illinois 60637, USA}
\affil[5]{Center for Fundamental Physics, Northwestern University, Evanston, Illinois 60208, USA}

\maketitle



\begin{abstract}
The application of silicon photomultiplier (SiPM) technology for weak-light detection at a single photon level has expanded thanks to its better photon detection efficiency in comparison to a conventional photomultiplier tube (PMT). 
SiPMs with large detection area have recently become commercially available, enabling applications where the photon flux is low both temporarily and spatially.
On the other hand, several drawbacks exist in the usage of SiPMs such as a higher dark count rate, many readout channels, slow response time, and optical crosstalk; therefore, users need to carefully consider the trade-offs.
This work presents a SiPM-embedded compact large-area photon detection module. Various techniques are adopted to overcome the disadvantages of SiPMs so that it can be generally utilized as an upgrade from a PMT. A simple cooling component and recently developed optical crosstalk suppression method are adopted to reduce the noise which is more serious for larger-area SiPMs.
A dedicated readout circuit increases the response frequency and reduces the number of readout channels.
We favorably compare this design with a conventional PMT and obtain both higher photon detection efficiency and larger-area acceptance.
\end{abstract}


\section{Introduction\label{sec:introduction}}
Optical sensors are essential devices in a wide range of scientific research. Various characteristics are required for different experimental systems.
In the case of advanced researches, weak-light detection at a single photon level is often required, such as fluorescence due to a relatively small number of atoms and molecules.
Among various highly sensitive photon sensors with single-photon detection capability, silicon photomultipliers (SiPMs) have recently emerged as high-performance alternative, with significant development and expanding list of applications\cite{Vilella2014,Zhao2017,Modi2019,Simon2019,Gundacker2020,Chung2022}.
Avalanche photodiodes (APDs) have high quantum efficiency but too much noise for low photon counts, photomultiplier tubes (PMTs) have low noise but low quantum efficiency. SiPMs fill in the gap.

A SiPM consists of an array of pixelated microcells of silicon APD. All microcells are reverse biased over the diode breakdown voltage. The excess of the bias voltage from the breakdown voltage which is called overvoltage determine the SiPM performance such as avalanche gain and dark count rate (DCR).  Since each microcell is sensitive to a single photon and has well-uniform avalanche gain among the microcells, the sum signal of all microcells contains photon number information.
In particular, SiPMs are attractive alternatives to PMTs in many cases due to their higher photon detection efficiency (PDE), lower operation voltage, and magnetic-field insensitiveness\cite{Wagatsuma2017,Piemonte2019,Giacomelli2019}.
Furthermore, SiPMs have extended the spectral range from near-infrared\cite{Tamura2018}, visible, to ultraviolet\cite{Acerbi2019}; currently it covers a similar spectral region as PMTs. 

Since only small SiPMs were available at the beginning of their development, the main applications were those that did not need a large sensitive area or those that required compact detectors. Examples include scintillation detectors with fine grain size\cite{Yeom2013,Lucchini2020}, or fiber readouts\cite{Izmaylov2010,Naito2016}.
Recent developments have expanded the sensitive area and now SiPMs as large as or even larger than PMTs are available. This enables us to use a SiPM for more general applications such as a replacement of a PMT. This is especially useful for spectroscopy of atoms and molecules where the emission signal tends to have a large etendue and it is difficult to secure sufficient geometrical acceptance with small sensors.
In comparison to PMTs, however, SiPMs have a number of disadvantages that need to be addressed: larger DCR\cite{Sul2013}, many readout channels, slower response time, and optical crosstalk (OCT)\cite{Nagy2014,Rosado2015}.

We develop a photon detector based on a large-area SiPM, overcoming the disadvantages of SiPMs, as a direct replacement for a conventional PMT for the ACME experiment\cite{Baron2014,Andreev2018}. The experiment aims to measure the permanent electric dipole moment of the electron\cite{Cesarotti2019}, by performing spin precession in a molecular beam, where the phase of the precesion is read out using laser-induced fluorescence (LIF) which emits photons at 512\,nm. 
The goal is to develop a SiPM based photon detector with higher visible PDE and improve photon number resolution. These lead to a higher efficiency in measuring fluorescence photons, increasing experiment sensitivity.

Here, we describe the design and performance of the developed module. We utilized a large-area array-type SiPM to maximize the sensitive area. To reduce the higher DCR, the SiPM sensor is cooled below the freezing point of water while the module casing is kept at room temperature. While small SiPM chips or arrays that equip similar cooling mechanisms are commercially available or reported\cite{Slenders2021}, our design enables good cooling performance for a large-area SiPM with a simplified cooling design.
Furthermore, custom optical filters are used to minimize the OCT, and a dedicated readout circuit has been developed to extend the response frequency and reduce the number of readout channels.

We did a comprehensive characterization of the modules and performed a test experiment using an actual molecular beam.
As well as the high PDE, the modules show good uniform single photoelectron gain among over 100 channels and noise levels almost consistent with the shot noise limit; all these parameters exceed those of conventional PMTs.
Sufficient cooling power is achieved using a simple mechanism which enables cooling the module down to -20$^\circ$C, where the DCR is less than 3\% of that at room temperature.
Finally, long-term stability of these parameters was confirmed in a high photon-yield application.

In this paper, we present the design of the module in section~\ref{sec:instruments}. Section~\ref{sec:characterization} reports the detailed characterization of the module on a test bench, then the molecular fluorescence measurement is described in Section~\ref{sec:beam}. Section~\ref{sec:summary} presents the summary and conclusion of the study.


\section{Instruments\label{sec:instruments}}
\subsection{Module design\label{sec:design}}

The SiPM array chip used is a commercial surface mount 16-channel array (Hamamatsu Photonics K.K. S13361-6075NE-04). It was selected because of its high PDE (45\% for wavelengths around 500\,nm) and the large sensor area.
The 16 channels are aligned as a $4 \times 4$ matrix with a gap between channels of 0.2\,mm, covering a total of $25 \times 25$\,mm square area. Each channel has $6 \times 6$\,mm square sensitive area that consists of 6336 $75 \times 75$\,$\mu$m microcells.
The SiPM has a DCR of typical $\sim$2\,M count per second (cps) for one channel at room temperature. This is quite higher than a PMT whose typical dark count is $\sim$kcps for a similar sensitive area\cite{Hamamatsu2017}. The OCT in this SiPM is typical 13\%.

\begin{figure}[ht]
\centering
\includegraphics[width=14cm, bb=0 0 1090 1060]{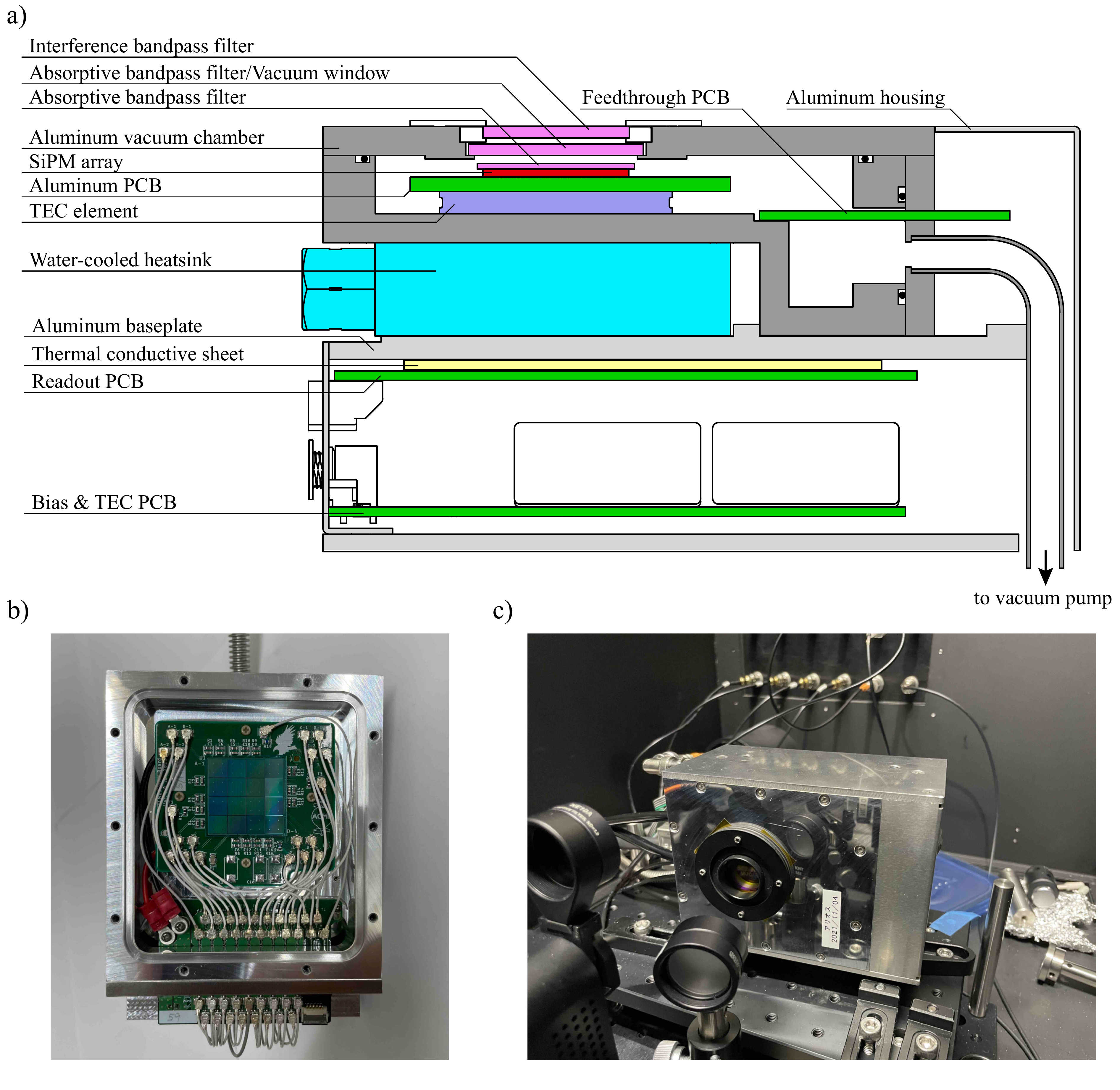}
\caption{a) Cutout view of the module. Electrical cables between PCBs are omitted. b) Picture of the inside the vacuum chamber. c) Front view of the mounted module. \label{fig:1}}
\end{figure}

A schematic and pictures of the module are shown in Fig.~\ref{fig:1}. To mitigate the DCR, the SiPM is housed in an aluminum vacuum chamber and is cooled down below the freezing point of water using a thermoelectric cooler element (TEC, Thorlabs, Inc. TECH4).
The SiPM is soldered onto a 2.0-mm-thick printed circuit board (PCB) made of aluminum instead of a usual glass epoxy to improve the thermal conductance between the SiPM and the TEC. Both surfaces of the TEC are attached to the PCB and the chamber wall using a thin layer of grease (Apiezon N). The SiPM and TEC heat load is transferred to a water-cooled heatsink on the outside of the chamber through the chamber wall itself. The water temperature is stabilized by a thermoelectric chiller to $\sim 20\,^\circ \mathrm{C}$ sufficiently high to avoid dew condensation around the chamber.
One negative temperature coefficient thermistor is mounted on the PCB and the PCB temperature is stabilized at a level of $0.01\,^\circ \mathrm{C}$ by a temperature controller (Arroyo instruments 7154-05-12). The PCB temperature is also monitored by two additional out-of-loop platinum temperature sensors on the PCB and a multimeter (Keithley DAQ6510 with 7710).

Three wavelength-selective optical filters are placed along the photon path to maximize the signal to noise ratio (SNR) and minimize OCT.
Due to crosstalk (including OCT and delayed crosstalk\cite{Nagy2014}) and afterpulse, a SiPM usually counts more than just the incident photons with a certain probability $p$. It effectively increases the single photoelectron gain from one pixel discharge and degrades the photon-counting resolution. The degradation is expressed as an excess noise factor. 
It is known that the gain is increased by a factor of $1+p$ and the excess noise factor is also $1+p$ by ignoring small $p^2$ terms\cite{Teich1986,Vinogradov2009}.
We have described the ideal filter configuration for the OCT suppression\cite{Masuda2021}, and we use a similar setup here.
Since the signal wavelength is 512\,nm in our case, we use bandpass filters (BPFs) whose transmission peaks are around that wavelength: an interference green BPF (Semrock FF01-520/70) whose center wavelength is 520\,nm and whose full width at half maximum is 70\,nm, a 2-mm-thick absorptive green BPF (SCHOTT BG39), and a 1-mm-thick absorptive green BPF (SCHOTT BG40) from the outside to the inside. The innermost filter is directly glued on the SiPM surface using transparent silicone potting (Momentive Performance Materials TSE3032). The resulting $p$ was $\sim0.1$ including crosstalk and afterpulse.

On the backside of the vacuum chamber, we placed bias and readout PCBs. Since the heat load of the readout circuit is not negligible ($\sim$5\,W), the readout circuit board is also attached by a thermal conductive sheet (3M 5580H) to the backplate which is directly in contact with the heatsink.
Signal and bias lines are connected via thin 50-$\Omega$ coaxial cables (Hirose, U.FL) from these PCBs to the SiPM PCB through the vacuum feedthrough PCB. All electronics are contained in an aluminum housing.

\subsection{Readout circuit\label{sec:readout}}
For effective use of large-area array type SiPMs, the readout circuit needs to be carefully considered for response speed and the number of readout channels.
In particular, the response speed would be an issue for a SiPM with a large sensor area because its capacitance is generally proportional to its surface area.
There are several methods to solve these issues. 

The number of readout channels can be reduced by electrically ganging several channels into one output. Suitable ways of ganging depend on requirements and constraints such as response speed, SNR, and cost. 
Passively connecting SiPMs in parallel is the simplest way but the response speed becomes slow because the net capacitance increases linearly with the number of SiPMs.
Series connection of SiPMs can reduce the net capacitance as the inverse of the number of connected SiPMs; however, it can only be used for pulse-like signal detection and not for DC-like signals due to the AC coupling. It also requires higher bias voltage proportional to the number of SiPMs, and the output signal charge is reduced.
Recently a hybrid connection readout has been proposed and used for some experiments\cite{Ieki2019,Chung2022}. The hybrid connection mitigates the high bias voltage in the series connection but still cannot be used for DC-like signals; thus, this method is effective for sparse pulse-like signals such as scintillation detectors. 
Another type of compensation method for the slow response speed is a pole-zero cancellation\cite{Gola2013} (PZC). This method uses a linear filter for shortening the slow tail by summing the original signal and the derivative pulse in time domain. Although this scheme needs careful parameter tuning to cancel the poles of the SiPM, it can shorten the slow tail component while keeping the DC-like signals intact.

\begin{figure}[ht]
\centering
\includegraphics[width=13cm, bb=0 0 786 422, clip]{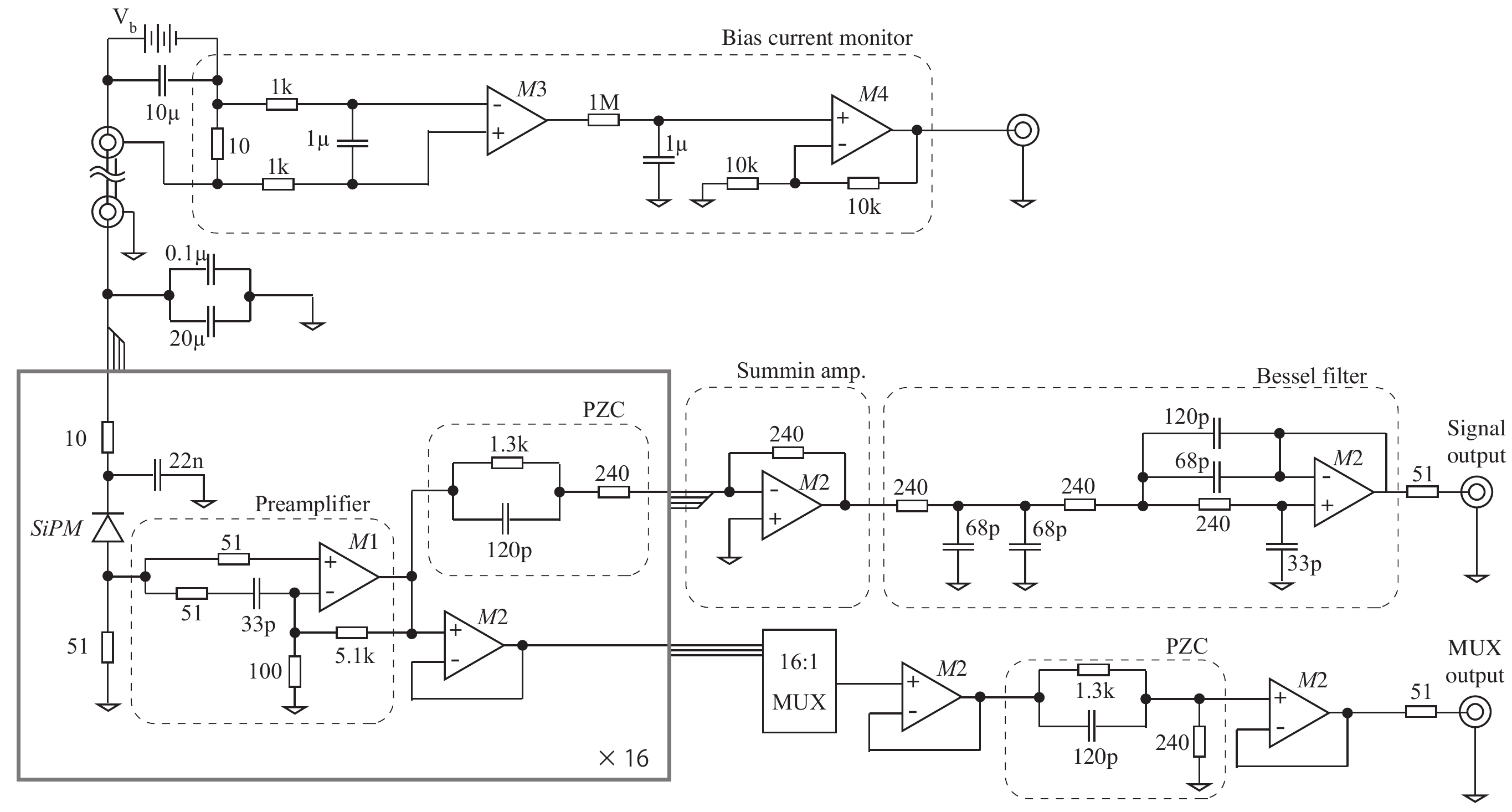}
\caption{Circuit diagram including a bias supply and readout. The power supply lines, bypass capacitors for operational amplifiers, and other less relevant components are omitted. $\mathrm{V_b}$: Bias supply; M1: LMH6629; M2: ADA4899; M3: AD620; M4: AD820. \label{fig:circuit}}
\end{figure}

We adopted the PZC technique to extend the response speed. We also developed an active summing circuit connected to the PZC directly. The circuit is shown in Fig.~\ref{fig:circuit}. The circuit is for one 16-channel array. Since the breakdown voltage of each channel is uniform enough, we simply use one DC supply (Keysight E36105B) to supply bias voltage for all channels. The individual channel output is transmitted to the backside amplifier board through coaxial cables and the vacuum feedthrough board, and then is simply terminated by a 51\,$\Omega$ resistor. The voltage signal across the termination resistor is buffered and sent to the PZC part.

The PZC outputs from all 16 channels are summed up by one operational amplifier to reduce the number of readout channels. The summed signal is also shaped by a following three-pole Bessel filter with a -3\,dB bandwidth of 5\,MHz to smooth the output signal so that the SNR is improved and the digitizing sampling rate can be reduced.

For diagnostic purposes, we extract signals from individual channels before the PZC part and send the one-channel signal through a multiplexer.
The bias current is monitored at the shunt resistor and an instrumentation amplifier.

\begin{figure}[ht]
\centering
\includegraphics[width=12cm, bb=0 0 431 310, clip]{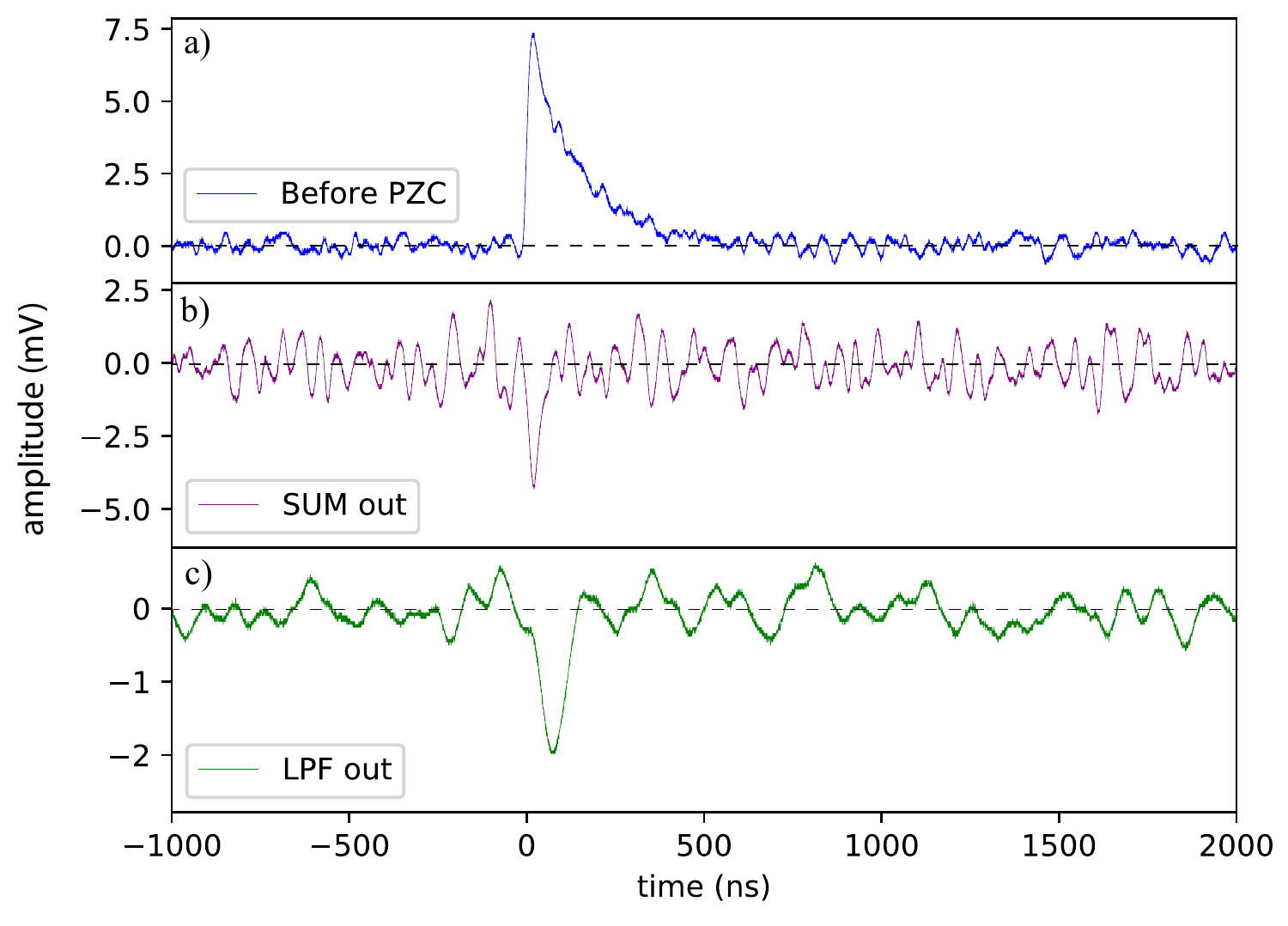}
\caption{Single photoelectron pulse shapes at three points of the readout circuit. Each offset is subtracted to zero. a) individual channel output before the PZC part. b) summed signal of 16 channels. c) Bessel filter output.\label{fig:singlepulse}}
\end{figure}

Single photoelectron pulse shapes at an overvoltage of 3.0\,V are shown in Fig.~\ref{fig:singlepulse}. The signal shape before the PZC part has a long tail whose decay constant is $\sim$100\,ns.
The PZC extends the -3\,dB bandwidth of the response frequency from 2\,MHz to 8\,MHz.
This value is better than the intrinsic response frequency of the SiPM. Since the SiPM's internal capacitance and resistance were measured in a separate setup to be 1.4\,nF and 30\,$\Omega$, respectively, it implies that even with an ideal transimpedance amplifier with zero input impedance, the resulting response frequency would be only 3.8\,MHz.
As shown in Fig.~\ref{fig:singlepulse}b, the single photoelectron pulse can also be clearly seen after the PZC even though 16 channels are summed and so the noise is also overlayed. The pulse shape is clearly shortened by the PZC; the decay constant is around 25\,ns.
The SNR is improved after the 5-MHz Bessel filter as shown in Fig.~\ref{fig:singlepulse}c.
This circuit enables a pulse counting method as well as a pulse integration method owing to the shortened pulse shape.
Since our required bandwidth is 5\,MHz, we tuned the PZC parameter as a bit faster than 5\,MHz and the pulse is shaped by the Bessel filter.
One can change the PZC parameter to make it much faster. Since it increases the noise level, one needs to tune based on the required response speed and noise level.

\section{Characterization\label{sec:characterization}}

In this section, we describe the basic characterization of the SiPM module on a test bench. We constructed 10 modules in total. With these modules, we measured 1) the single photoelectron gain as an integral area of a single photoelectron pulse, 2) the temperature dependence of the breakdown voltage, 3) the temperature dependence of the DCR, 4) the linearity of the output signal, and 5) the excess noise factor and the relative PDE. We also measured the long term stability of the SiPM module while continuously injecting photons. The module was set in a light-tight box to block ambient light, and a thermo-stabilized 515\,nm laser diode (OSRAM PLT5) was used as a light source for the linearity, excess noise factor, and long term stability measurements. A scroll pump was used for keeping vacuum inside the SiPM housing around 1\,Pa, the water temperature of the heatsink was controlled to be 20$^\circ$C, and the TEC was used in tandem with a PID controller to keep the thermally isolated parts to be a nominal temperature of -15$^\circ$C, unless otherwise stated. In the following measurements, a 2-channel oscilloscope (NI PXIe-5164) was used to record waveforms with a sampling rate of 100\,MHz and a -3\,dB bandwidth of 20\,MHz. The breakdown voltage of each SiPM array was measured before the characterization test. We applied 3.0\,V overvoltage, which is the recommended value by the manufacture\cite{Hamamatsu2021} to obtain enough gain factor while keeping the affects from the afterpulse, OCT, and DCR acceptable.\\

\noindent {\bf Single photoelectron gain}\\
Dark count noise was recorded to obtain a single photoelectron waveform of the summed and filtered signal.
A single-photon area is defined as an integral of the waveform (Fig.~\ref{fig:singlepulse}(c)) between -100\,ns and 200\,ns from the leading edge, and it is obtained in a unit of pV$\cdot$s. A single-photon area distribution is obtained from a 12.5\,ms trace, where the DCR at -15$^\circ$C is roughly 20\,kcps for each channel. The typical value is measured to be around 175\,pVs for the summed signal.

\begin{figure}[h]
\centering
\begin{minipage}{0.47\textwidth}
\centering
\includegraphics[width=6.3cm, bb=0 0 548 402]{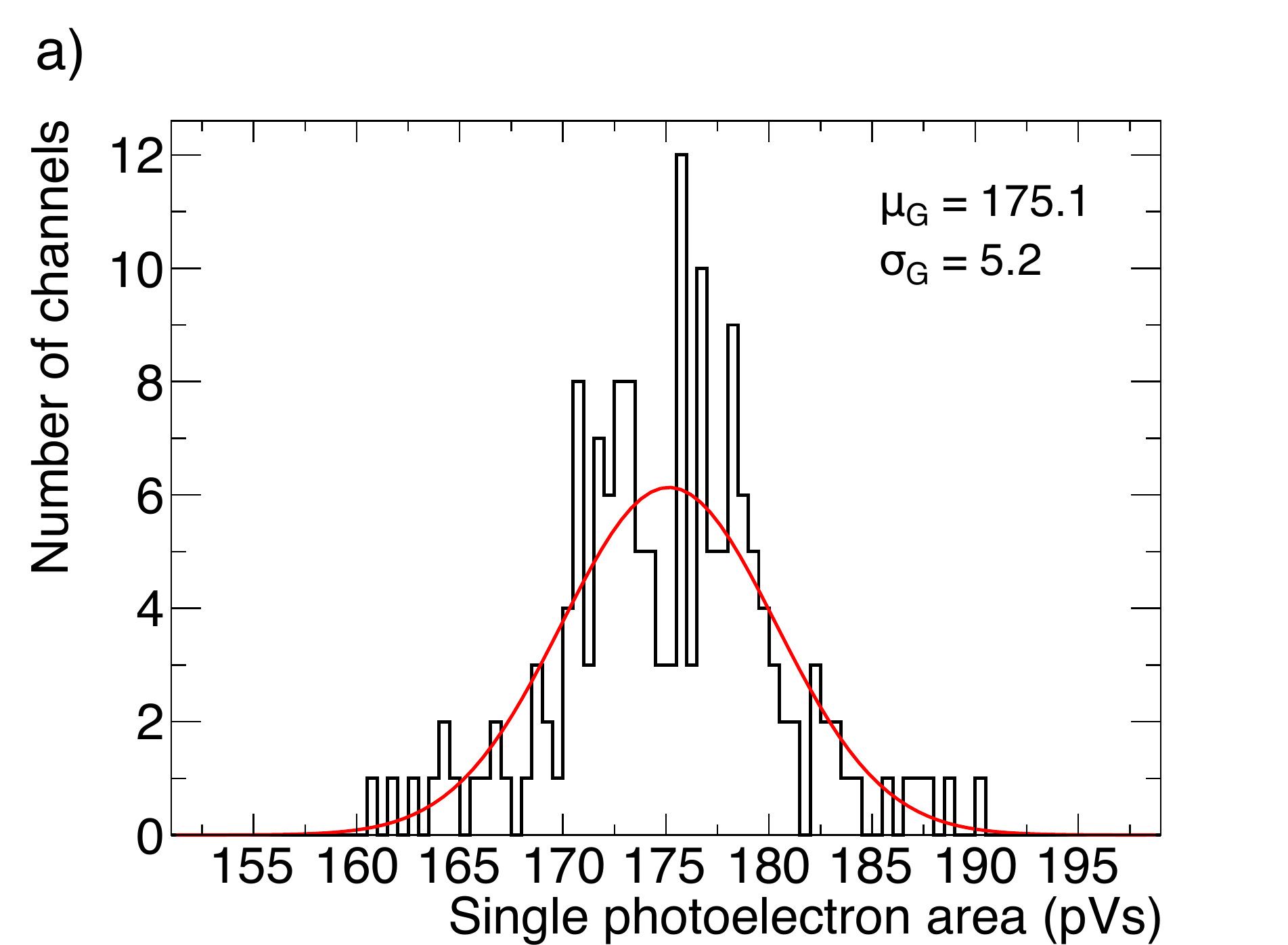}
\end{minipage}
\begin{minipage}{0.47\textwidth}
\centering
\includegraphics[width=6.3cm, bb=0 0 552 402]{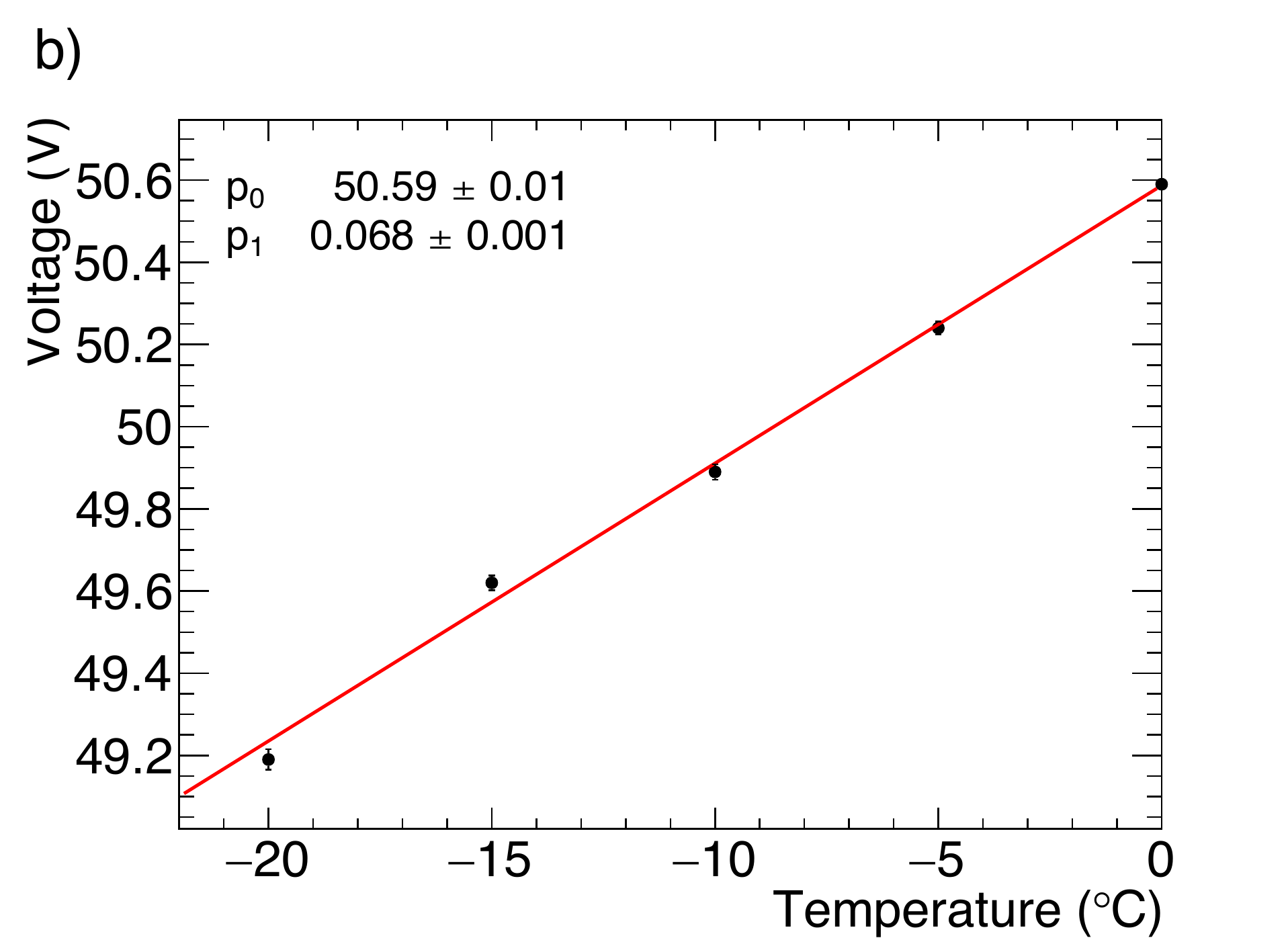}
\end{minipage}
\caption{a) Single photoelectron area of all the channels. The values $\mu_G$ and $\sigma_G$ are the mean and the standard deviation obtained by a Gaussian fitting, respectively. The effect of the deviation to the photon counting resolution is less than 0.1\%. b) Relation between temperature at the SiPM-mounted PCB and the breakdown voltage fit by a linear function, where the parameters $p_0$ and $p_1$ are the intercept and the slope, respectively.\label{fig:SPArea}}
\end{figure}

We also measured the single-photon area of an individual SiPM channel in a similar manner using the multiplexer output. A Gaussian fitting of the single-photon area distribution gives the mean value. Figure \ref{fig:SPArea}(a) shows a distribution of the single-photon area of each channel. Although the SiPM module does not have a bias adjustment system for each channel of the SiPM array, the deviation of the single-photon area is only 3\% owing to very similar breakdown voltages among channels. The effect of the deviation to the photon counting resolution is less than 0.1\%, and it is negligible compared with the effects of the crosstalk and the afterpulse ($\sim$7\%), which are expressed in the measurement of the excess noise factor.\\

\noindent {\bf Temperature dependence of the breakdown voltage}\\
 Figure \ref{fig:SPArea}(b) shows the measured breakdown voltage as a function of temperature at the SiPM-mounted PCB, where a clear linear relationship is observed. Since the breakdown voltage is known to be linear to the SiPM chip temperature\cite{Hamamatsu2021}, this result suggests that the SiPM chip is also cooled down as a linear function of the temperature at the PCB down to -20$^\circ$C.

\begin{figure}[h]
\centering
\begin{minipage}{0.47\textwidth}
\centering
\includegraphics[width=6.3cm, bb=0 0 548 402]{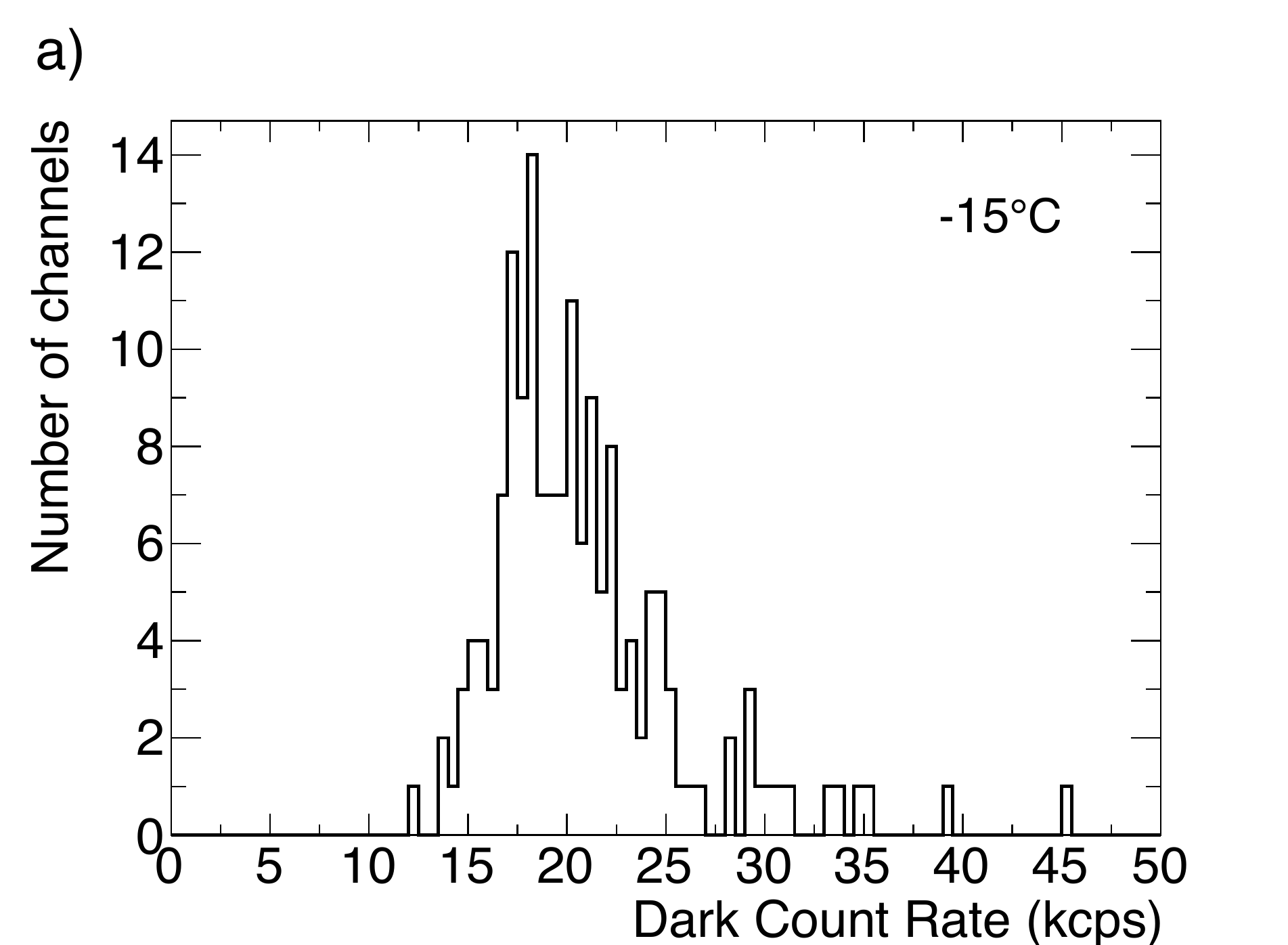}
\end{minipage}
\begin{minipage}{0.47\textwidth}
\centering
\includegraphics[width=6.3cm, bb=0 0 548 402]{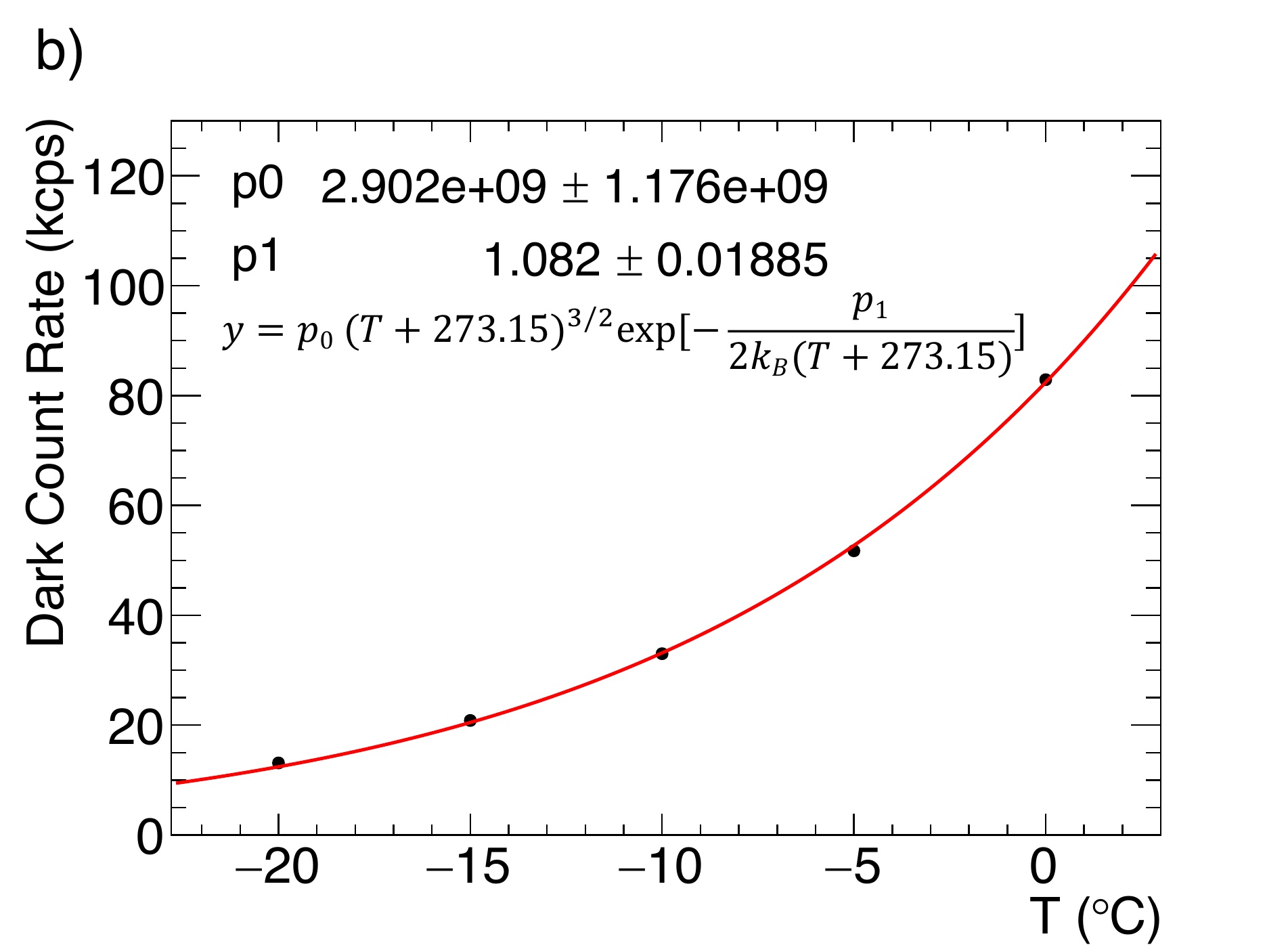}
\end{minipage}
\caption{a) Histogram showing DCR of all the channels at -15$^\circ$C. b) Temperature dependency of the DCR. The DCR at low temperature is suppressed as expected.\label{fig:DCR}}
\end{figure}

\noindent {\bf Dark count rate}\\
The DCR of the individual channels is measured by changing the TEC operation temperature every 5$^\circ$C between -20$^\circ$C and 0$^\circ$C. A 12.5\,ms trace was recorded for each channel using the multiplexer output. We set the threshold as 1.5\,mV over the baseline, and peaks exceeding the threshold are counted as dark counts. Figure \ref{fig:DCR} shows the distribution of the DCR of all channels at -15$^\circ$C and the temperature dependency of the DCR, which can be approximated by an exponential function\cite{Hamamatsu2021}. In this figure, the mean value of the DCR of all the channels is plotted. The DCR at -15$^\circ$C is suppressed to 3\% of that at room temperature, which is extrapolated using a function shown in Fig. \ref{fig:DCR}.\\

\begin{figure}[h]
\centering
\begin{minipage}{0.47\textwidth}
\centering
\includegraphics[width=6.3cm, bb=0 0 548 402]{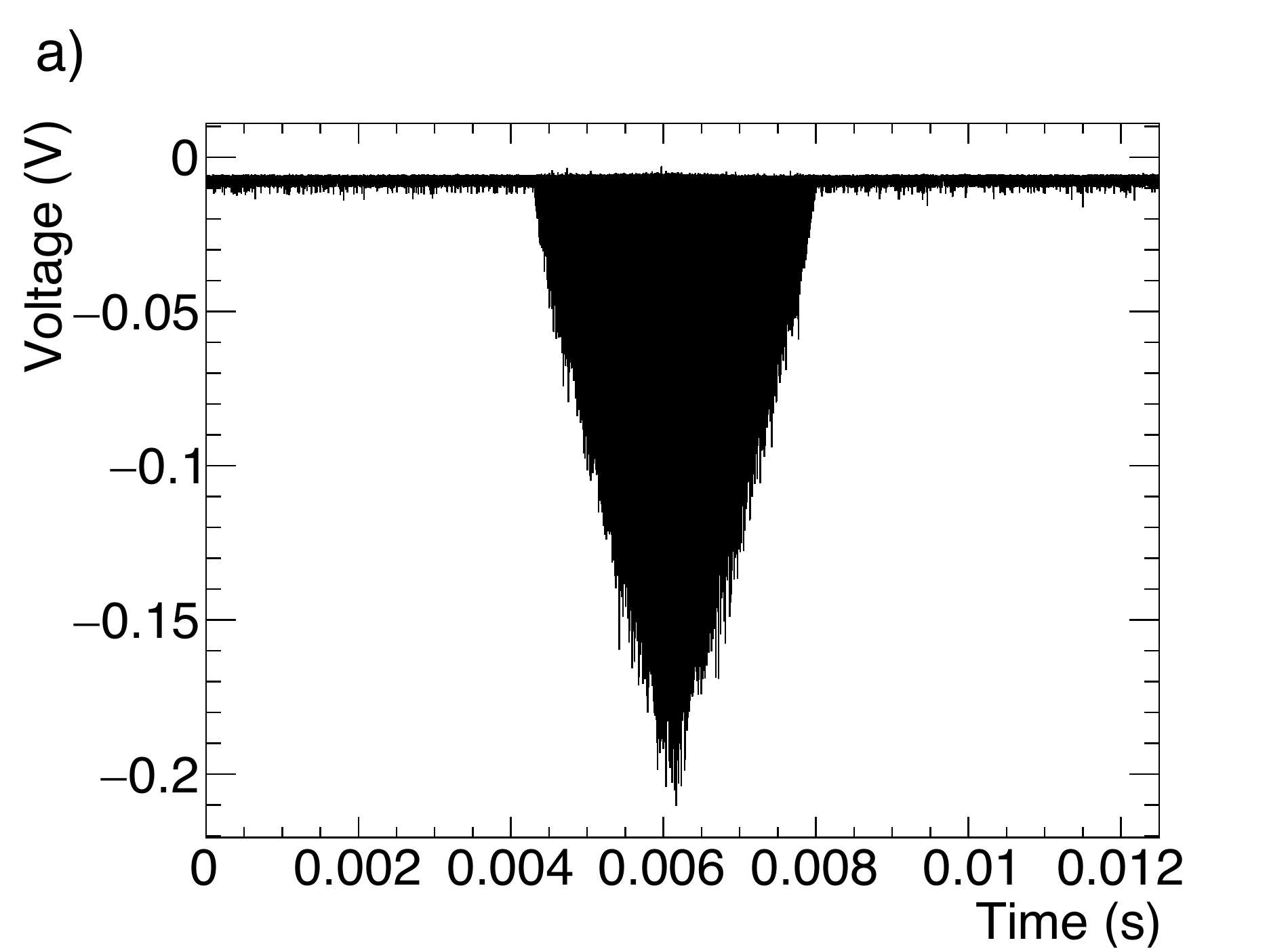}
\end{minipage}
\begin{minipage}{0.47\textwidth}
\centering
\includegraphics[width=6.3cm, bb=0 0 548 402]{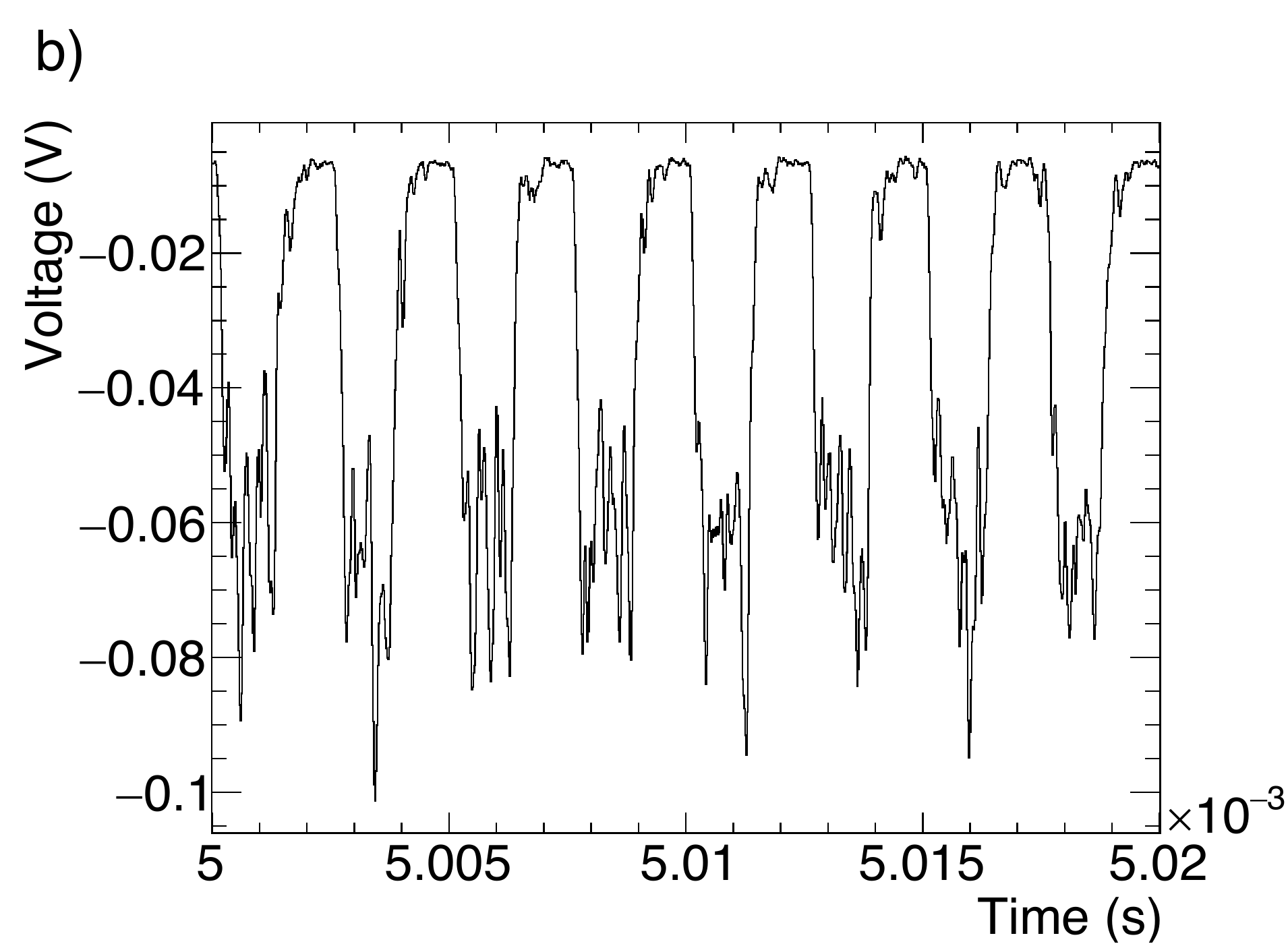}
\end{minipage}
\caption{a) Overall envelope of the square pulses created by a 515\,nm laser diode. b) Zoomed view of the pulse structure corresponding to the range from 5.00\,ms to 5.02\,ms in the left figure.\label{fig:Triangle_trace}}
\end{figure}

\noindent {\bf Signal linearity}\\
The output signal linearity was measured using the laser diode. Pulsed light was injected repeatedly by modulating the laser current using an arbitrary waveform generator (siglent SDG2042X) to measure the deviation of the measured number of photons. The light from the laser diode is reduced by more than three orders of magnitude and adjusted using ND filters. We injected a 400\,kHz square pulse whose envelope is a triangle as shown in Fig. \ref{fig:Triangle_trace}. Each 2.5\,$\mu$s cycle is called a bin, and the number of photons in each bin is measured. The number of photons in a bin was around 8000 photons at the peak of the triangle envelope.

\begin{figure}[h]
\centering
\includegraphics[width=9cm, bb=0 0 548 402]{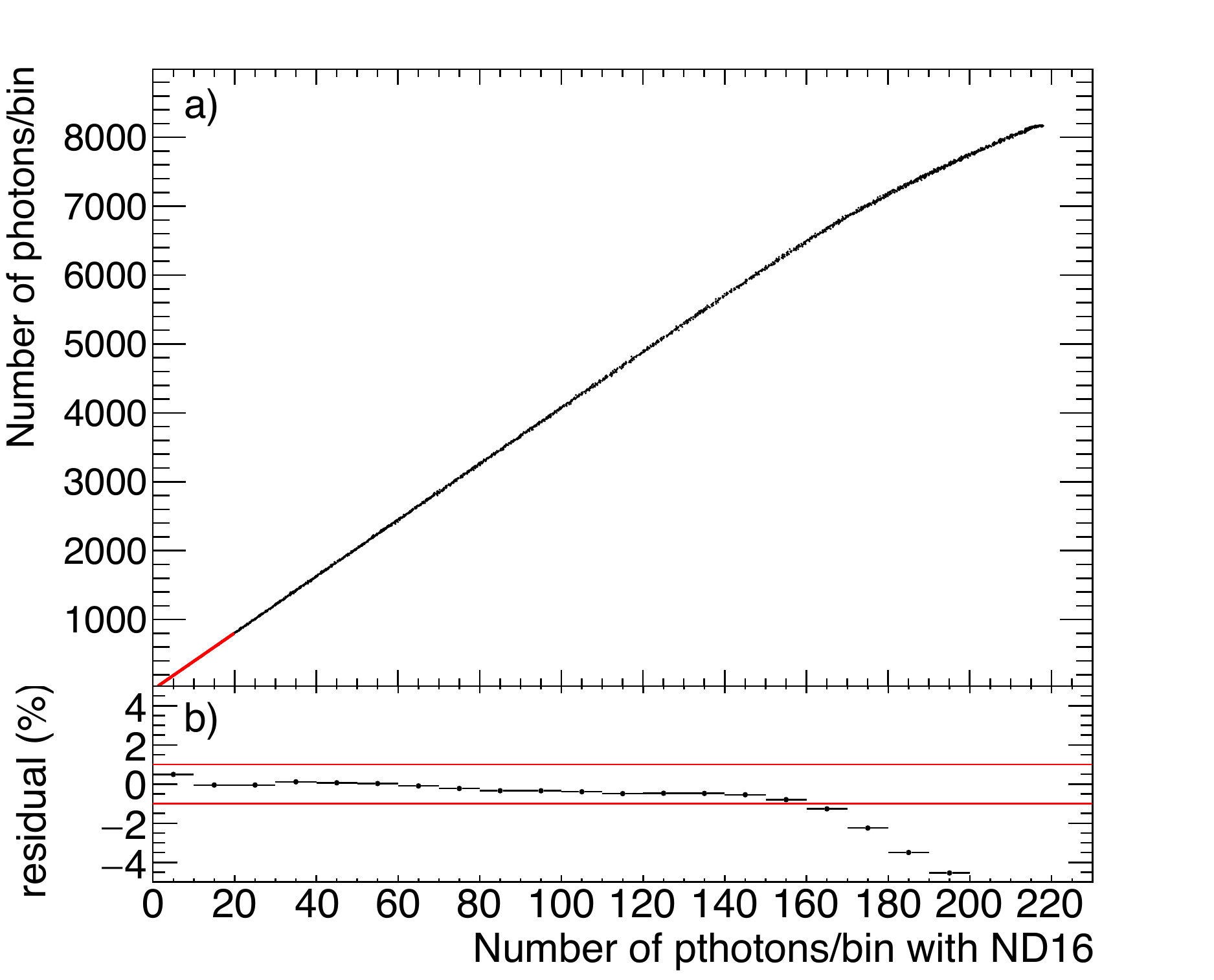}
\caption{a) Comparison of the number of photons at each bin, with and without an ND16 filter in front of the SiPM module. b) Residual from a linear fitting. The linearity is within $\pm1\%$ up to around 6000 photons.\label{fig:linearity}}
\end{figure}

Figure \ref{fig:linearity} shows a comparison of the number of photons with and without an ND16 filter in front of the SiPM module. Each point in the plot corresponds to the number of photons at each bin. A linear fit is applied to a low yield region, and the residual from the fit result is shown at the bottom of Fig. \ref{fig:linearity}. The linearity is within $\pm1\%$ up to around 6000 photons, and we observe a saturation of the output signal above it. This saturation is not due to the SiPM but comes from the saturation of gain in the readout circuit. Therefore, we can reach better linearity by reducing the readout circuit gain, if necessary, while the linearity is good enough for our use since the expected number of photons is around 1000 at the maximum.

\noindent {\bf Excess noise factor and the relative PDE}\\
We can measure the excess noise factor using the same setup with the measurement of the signal linearity. When the injected light has a Poisson distribution, it is known that the Fano factor of the number of measured photons taking into account the effective gain also becomes $1+p$, which is the same expression as the excess noise factor\cite{Vinogradov2009,Vinogradov2016}. Here, the effective gain is calculated by multiplying the single-photoelectron gain by $1+p$, and we assumed $p=0.1$ in the following measurement. The Fano factor ($F$) can be measured as $F=\sigma_N^2/\mu_N$, where $\sigma_N$ and $\mu_N$ are the standard deviation and mean value of the number of photons obtained from 1000-shot waveforms (Fig. \ref{fig:Triangle_trace}), respectively. The number of photons is calculated by dividing the waveform integral by the effective gain at each bin. The number of photons in a bin was set to be 1000 photons at the peak. 

\begin{figure}[h]
\centering
\includegraphics[width=8cm, bb=0 0 556 399]{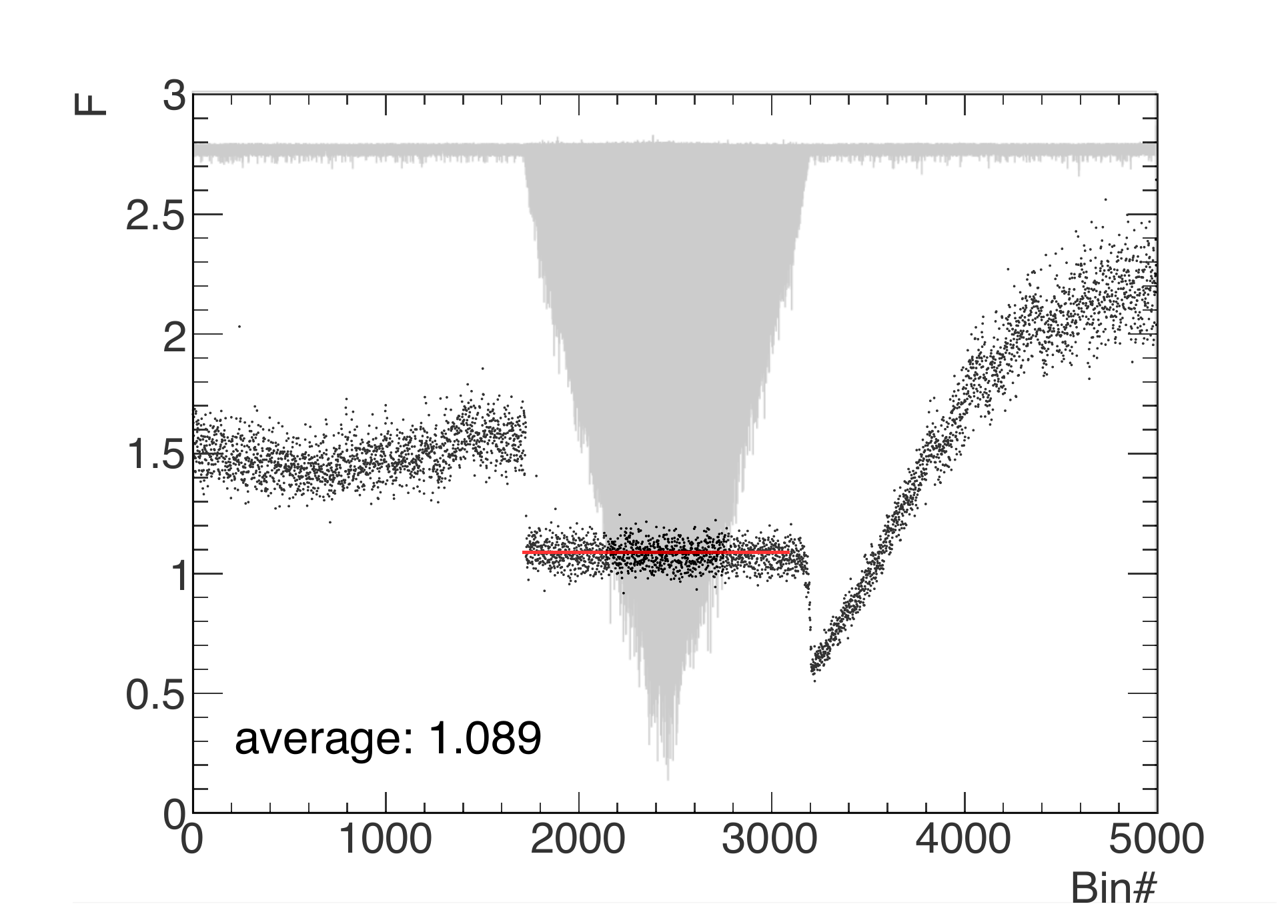}
\caption{Example of the Fano factor at each bin. The average value for the 10 modules is $F\sim 1.07$. The gray plot shows the signal envelope. The Fano factor in the bins without light injection are not estimated properly because they are strongly affected by fluctuation of the signal baseline.\label{fig:ExcessNoise_overlay}}
\end{figure}

Figure \ref{fig:ExcessNoise_overlay} shows an example of the measured Fano factor of a module. The average value of $F$ for the 10 modules is $F\sim 1.07$. This result is consistent with the assumption of $p\sim0.1$, and the excess noise factor obtained from this measurement is smaller than those of typical PMTs, $\sim$1.2\cite{headon}

Since we injected a certain number of photons repeatedly in the above measurement, the relative PDE can be obtained by comparing the number of measured photons between the 10 modules. We found that the relative photon detection efficiencies are consistent within $\pm$5\%.\\

\begin{figure}[h]
\centering
\includegraphics[width=12cm, bb=0 0 540 390]{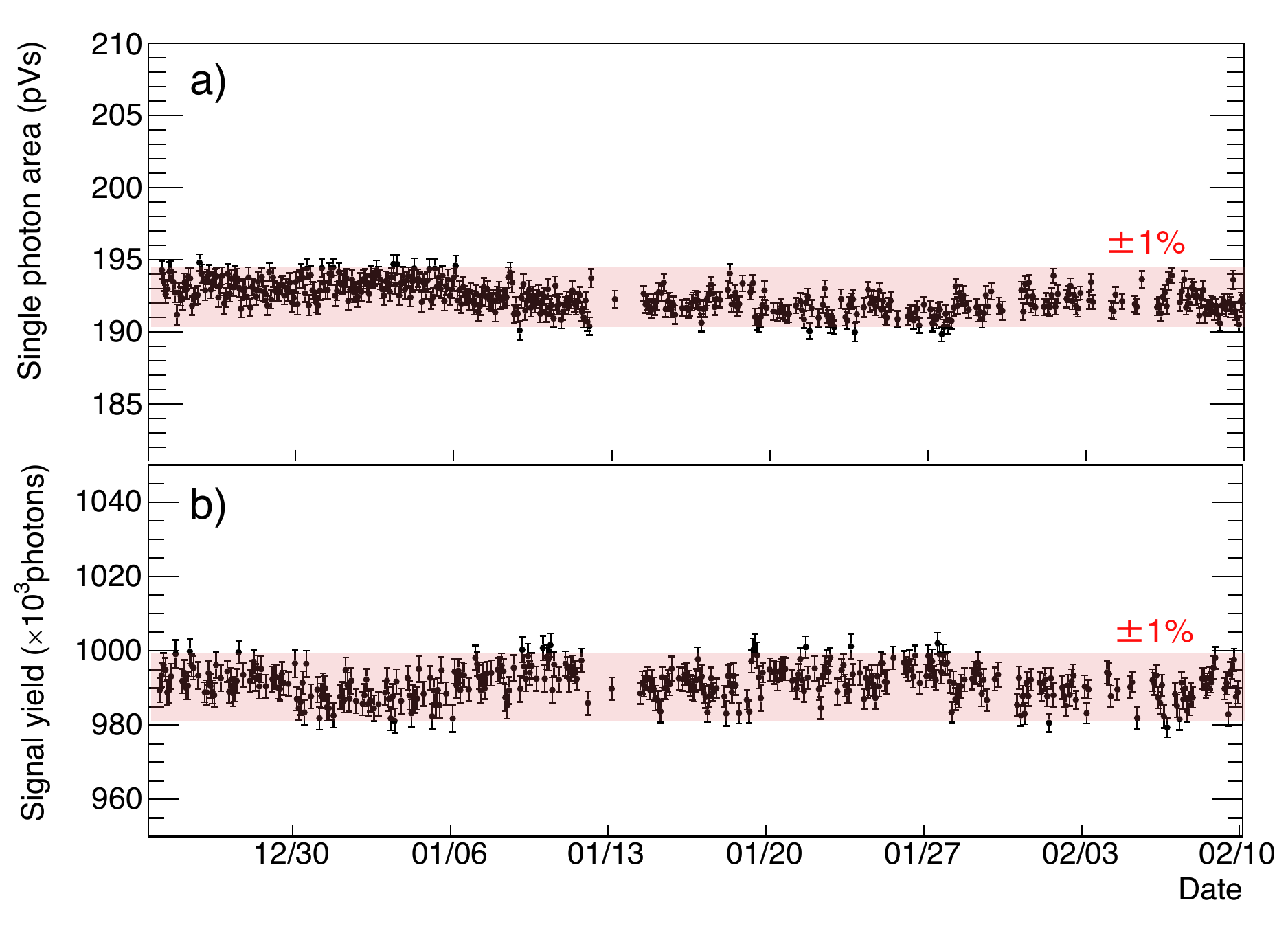}
\caption{a) Stability of the single photoelectron gain during the long term operation. b) Signal yield stability. Each red band corresponds to a $\pm1\%$ region, and they are stable within 1\% over seven weeks. \label{fig:long}}
\end{figure}

\noindent {\bf Long term operation}\\
Long-term measurements of detector performance were carried out over seven weeks with short breaks due to other tests. The purpose of this operation is to test the stability of the single photoelectron gain and the PDE of the SiPM module. In this measurement, we used a 4-channel oscilloscope (Tektronix MSO64) with a sampling rate of 125MHz and a -3\,dB bandwidth of 20\,MHz, and a 10\,ms waveform trace was recorded every one minute.  The number of photons in a bin at the peak was set to 1000. Figure \ref{fig:long} shows the single photoelectron area and the signal yield during the long term operation. The signal yield is obtained as the total number of photoelectrons in 10\,ms. They are stable within 1\% over seven weeks, while the integrated number of detected photons is about 1.7$\times10^{14}$. In the case of PMTs, output current could change by around 10\% after detecting this amount of photons due to damage of the last dinode\cite{Hamamatsu2017}, while the deviation of the SiPM module efficiency remained within $\pm$1\% for the entire duration of the test.
\\


\section{Molecular fluorescence measurement\label{sec:beam}}
In this section, we show an example of the SiPM module application to LIF measurements of molecules. Specifically, we used the setup of the ACME II experiment\cite{Andreev2018} and observed the fluorescence of thorium monoxide (ThO) molecules.
ThOs initially prepared in the \textit{Q} state were excited with a laser resonant with the \textit{Q-I} transition. We use the detector to count fluorescence photons with a wavelength of 512\,nm from the decay of the \textit{I} to \textit{X} state.

\begin{figure}[ht]
\centering
\includegraphics[width=11cm, bb=0 0 874 280, clip]{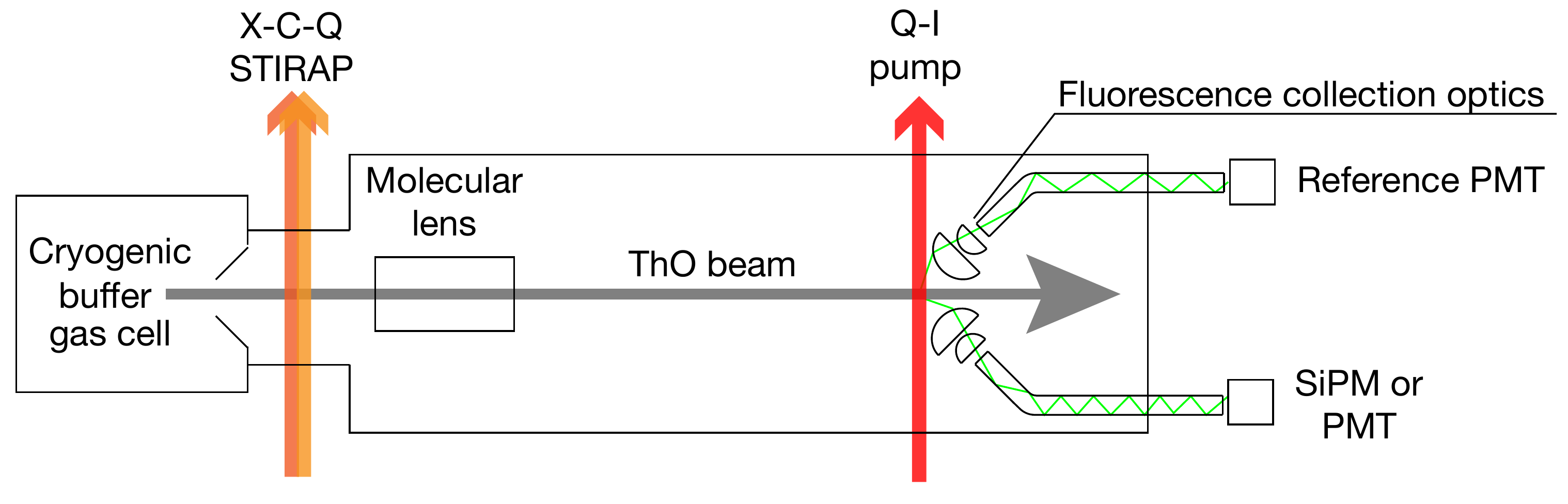}
\caption{Schematic view of the ThO beam line setup. ThO molecules are produced in the cryogenic buffer gas cell at the leftmost part of the beam line, then the molecules go rightward. \label{fig:beamline}}
\end{figure}

The experimental setup is shown in Fig.~\ref{fig:beamline}. The ThO molecules were produced in the leftmost chamber. A ThO$_2$ pellet was mounted inside the chamber. A pulsed YAG laser ablated the pellet and ThO molecule gas was formed. The gas was cooled down by 17\,K neon buffer gas, forming a cryogenic molecular beam mostly in the ground (\textit{X}) electronic state\cite{Hutzler2012}. The molecules in the beam were then excited to the \textit{Q} state using Stimulated Raman Adiabatic Passage (STIRAP) via \textit{X-C-Q} transitions \cite{Wu2020} and were collimated by an electrostatic molecular lens to the readout region downstream\cite{Wu2022}.
The molecules in the \textit{Q} state were resonantly excited to the \textit{I} state by a 746\,nm continuous-wave laser (M Squared SolsTis), which decays quickly (lifetime $115 \pm 4$\,ns\cite{Kokkin2014}) to the ground state \textit{X}, emitting 512\,nm photons. The photons were collected by an optical lens doublet\cite{PANDA2018} and guided to the SiPM or PMT (Hamamatsu photonics K.K. R7600-300U) through a 16-mm-diameter cylindrical quartz light pipe.

We replaced the SiPM with the PMT to compare the absolute signal detection rate. The PMT had the same interference bandpass filter in front of the window.
The molecular beam intensity was monitored by another equivalent PMT set which was mounted on another light pipe.
The PMT output was amplified by a factor of 25 through a preamplifier (Stanford Research System SR445A) twice and was shaped by a 5\,MHz low pass filter. Both signals were digitized by an oscilloscope (NI PXIe-5162) with a 100\,MHz sampling rate.
The SiPM was operated at an overvoltage of 3.0\,V at the temperature of -20\,$^\circ$C. The applied high voltage of the PMT was -800\,V.
The gain of both the SiPM and the PMT were calibrated just before the measurement. 

\begin{figure}[ht]
\centering
\includegraphics[width=8cm, bb=0 0 531 343, clip]{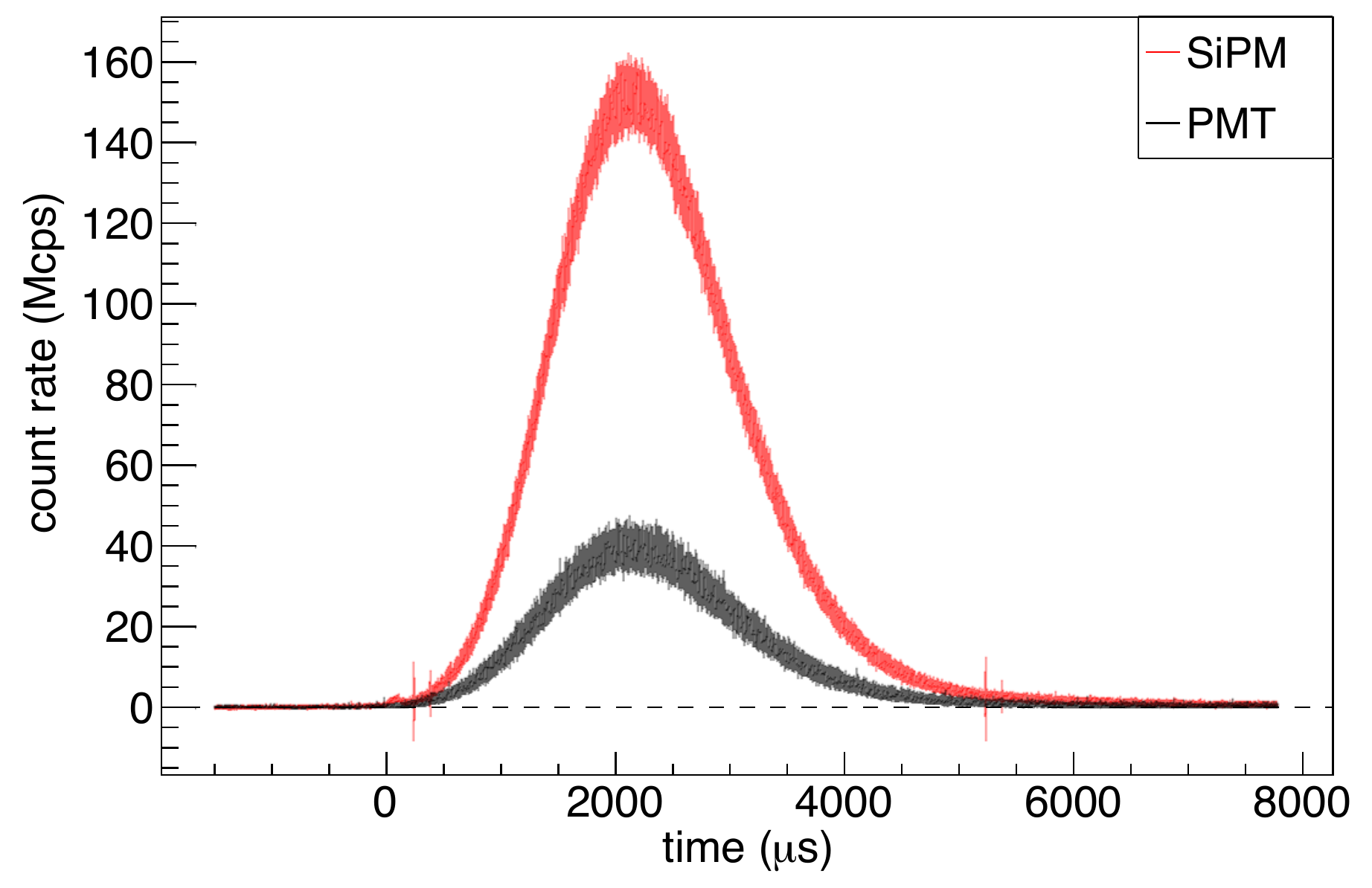}
\caption{Molecular signal count rate. The PMT count rate is scaled by the beam intensity monitor. The spike noises at 250\,$\mu$s and 5250\,$\mu$s are likely correlated to 200\,Hz flashlamp operation in the ablation YAG laser.\label{fig:molsignal}}
\end{figure}

The signal count rates are shown in Fig.~\ref{fig:molsignal}. Both signals were averaged 100 shots and smoothed with a sliding time window of 100\,ns. 
The module has a higher detection rate than the PMT by a factor of more than 3. Note that the PMT detection rate was scaled by the beam intensity monitor so that both detection rates can be directly compared. The improvement of the photo detection efficiency itself is at a level of a factor of $\sim$2.3. The PMT has an effective area of 18 $\times$ 18\,mm square, and our SiPM chip has an effective area of 24 $\times$ 24\,mm square. The remaining factor seems to be covered by the effective area and alignment, since the SiPM and PMT surfaces were placed slightly far from the lightpipe end.
This demonstrates an important improvement for the next generation of the ACME experiment. 

\section{Summary\label{sec:summary}}
The recent development of a large-area silicon photomultiplier array enables an upgrade from a conventional photomultiplier tube to improve the photon detection efficiency. We developed a SiPM-array embedded photon detector that mitigates the disadvantages of SiPMs.
It uses cooling of the SiPM array to suppress DCR. The readout circuit reduces the number of channels from 16 to 1 with low enough baseline noise to distinguish single photoelectron pulses. It also provides frequency response from DC to above the intrinsic bandwidth of the SiPM.

We performed a comprehensive characterization of the modules.
The uniformity of the single photoelectron gain and PDE are better than 3\% and 5\%, respectively. The module has a good cooling capability and the resultant DCR at -15$^\circ$C is successfully suppressed to 3\% of that at room temperature. We also confirmed that the output signal linearity is within $\pm$1\% up to around 6000 photons per $\sim$1\,$\mu$s. The excess noise factor was measured to be $F\sim 1.07$. Finally, the long-term stability test shows the single photoelectron gain and the PDE were stable within 1\% over seven weeks.

These studies demonstrate that a replacing PMT with large-area SiPM can be done in a robust way that takes advantage of the increased PDE for general applications such as spectroscopy of atoms and molecules as shown in this research.

\section*{Funding}
Okayama University (RECTOR program); Japan Society for the Promotion of Science (KAKENHI JP20KK0068, JP21H01113, JP21J01252); Matsuo Foundation; National Science Foundation; Gordon and Betty Moore Foundation; Alfred P. Sloan Foundation.

\section*{Acknowledgments}
We would like to thank Z. Lasner for initial considerations of using SiPMs and discussion about the characteristics, and N.R. Hutzler for the suggestion of the optical crosstalk suppression.
We also appreciate T. Kobayakawa, D. Lascar, Z. Han, P. Hu, and S. Liu for helping the development.

\bibliographystyle{unsrt}
\bibliography{SiPMForACME}

\end{document}